\documentclass[%
 reprint,
 amsmath,amssymb,
 aps,prl,longbibliography,superscriptaddress
]{revtex4-2}
\usepackage[utf8]{inputenc}
\usepackage{newunicodechar,textgreek}

\usepackage{graphicx}       
\usepackage{dcolumn}        
\usepackage{bm}             
\usepackage{siunitx}        
\DeclareSIUnit{\gauss}{G}   
\usepackage{physics}        
\usepackage[english]{babel}

\usepackage{xparse}         
\usepackage{color}          
\usepackage{xcolor}         
\usepackage{comment}        
\usepackage{import}         
\usepackage{hyperref}       
\usepackage[normalem]{ulem}

\renewcommand{\selectlanguage}[1]{}

\begin{document}

\preprint{APS/123-QED}

\title{Probing topological edge states in a molecular synthetic dimension}

\author{Adarsh P. Raghuram}
\affiliation{Department of Physics, Durham University, South Road, Durham DH1 3LE, United Kingdom.}

\author{Francesca M. Blondell}
\affiliation{Department of Physics, Durham University, South Road, Durham DH1 3LE, United Kingdom.}

\author{Jonathan M. Mortlock}
\affiliation{Department of Physics, Durham University, South Road, Durham DH1 3LE, United Kingdom.}

\author{Benjamin~P.~Maddox}
\affiliation{Department of Physics, Durham University, South Road, Durham DH1 3LE, United Kingdom.}

\author{Sohail~Dasgupta}
\affiliation{
School of Physics and Astronomy, University of Birmingham, Edgbaston, Birmingham B15 2TT, United Kingdom.}

\author{\mbox{Holly A. J. Middleton-Spencer}} 
\affiliation{
School of Physics and Astronomy, University of Birmingham, Edgbaston, Birmingham B15 2TT, United Kingdom.}

\author{\mbox{Kaden R. A. Hazzard}} 
\affiliation{
Department of Physics and Astronomy, Rice University, Houston, Texas 77005, USA.}
\affiliation{
Smalley-Curl Institute, Rice University, Houston, Texas 77005, USA.}

\author{Hannah M. Price} 
\affiliation{
School of Physics and Astronomy, University of Birmingham, Edgbaston, Birmingham B15 2TT, United Kingdom.}

\author{Philip D. Gregory}
\affiliation{Department of Physics, Durham University, South Road, Durham DH1 3LE, United Kingdom.}

\author{Simon L. Cornish}
\affiliation{Department of Physics, Durham University, South Road, Durham DH1 3LE, United Kingdom.}

\date{\today}
\begin{abstract}
Engineering synthetic dimensions, where the physics of additional spatial dimensions is simulated within the internal states of a quantum system, allows the realisation of phenomena not otherwise accessible in experiments. Ultracold ground-state polar molecules are an ideal platform to encode synthetic dimensions, offering access to large Hilbert spaces of long-lived internal states associated with the rotational and hyperfine degrees of freedom, that can be coupled together with microwave fields to simulate tunnelling. Here, to benchmark the advantages of ultracold molecules, we encode a 1D synthetic lattice in the rotational states of ultracold RbCs molecules and use it to investigate the well-known Su-Schrieffer-Heeger (SSH) model, a minimal model displaying topological properties. To probe the system, we perform spectroscopy using an auxiliary rotational state and study the time dynamics after deterministic state preparation. We demonstrate long coherence times, typically $\sim500$ times the lattice tunnelling period, even for a synthetic lattice using 8 rotational states.
Observations of dynamics at long times with full site-resolved readout of the synthetic dimension allow us to test the effects of chiral and non-chiral perturbations on the topologically protected edge states.
Our work lays the foundation for further quantum simulations using the rich internal structure of molecules, including dipolar string phases in interacting samples of molecules, and adiabatic state preparation of many-body Hamiltonians.

\end{abstract}
\maketitle

Ultracold polar molecules combine rich internal structures with access to tunable, long-range and anisotropic dipolar interactions. These properties have led to many proposed applications in quantum computing~\cite{demilleQuantumComputationTrapped2002} and quantum simulation~\cite{gorshkovTunableSuperfluidityQuantum2011,cornishQuantumComputationQuantum2024a,micheliToolboxLatticespinModels2006}. Most proposals utilise the rotational states of the molecules to encode qubits~\cite{gregoryRobustStorageQubits2021,picardEntanglementISWAPGate2025,yelinSchemesRobustQuantum2006,niDipolarExchangeQuantum2018}, qudits~\cite{sawantUltracoldPolarMolecules2020}, or pseudospins for the simulation of quantum spin models~\cite{cornishQuantumComputationQuantum2024,micheliToolboxLatticespinModels2006,barnettQuantumMagnetismMulticomponent2006}. However, an alternative approach, proposed by Sundar~\emph{et~al.}~\cite{sundarSyntheticDimensionsUltracold2018,sundarStringsUltracoldMolecules2019}, is to exploit the rotational states of polar molecules to realise \emph{synthetic dimensions}. 

Synthetic dimensions describe additional degrees of freedom encoded in the discrete internal states of a system used to simulate lattice sites in an additional spatial dimension~\cite{ozawaTopologicalQuantumMatter2019,boadaQuantumSimulationExtra2012,yu_comprehensive_2025}. Coherent coupling between the internal states is then interpreted as tunnelling in the synthetic lattice. The approach has several advantages: (i)~It allows the simulation of higher-dimensional systems. For example, atoms confined in 1D or 2D real-space lattices can be used to simulate higher dimensional physics~\cite{viebahnMatterWaveDiffractionQuasicrystalline2019,priceFourDimensionalQuantumHall2015,bouhiron_realization_2024}. (ii)~Engineering site-specific complex tunnelling amplitudes in the synthetic dimension is straightforwardly achieved by controlling the strength and phase of the coupling fields. (iii)~Changing the geometry of the synthetic dimension simply requires changing the coupling fields, rather than implementing a bespoke optical lattice. (iv)~Resolving single sites in the synthetic dimension is readily achieved using state-selective imaging techniques. These advantages have been used to study a range of phenomena including synthetic gauge fields~\cite{celiSyntheticGaugeFields2014}, Anderson localisation~\cite{meierObservationTopologicalAnderson2018}, and Thouless pumping~\cite{trautmannRealizationTopologicalThouless2024}. 

Synthetic dimensions have been implemented in experiments with photonic systems~\cite{linPhotonicWeylPoint2016,duttSinglePhotonicCavity2020,yuanSyntheticDimensionPhotonics2018,liDirectExtractionTopological2023,wangVersatilePhotonicFrequency2025} and cold atomic systems exploiting atomic hyperfine states~\cite{manciniObservationChiralEdge2015,stuhlVisualizingEdgeStates2015,kolkowitzSpinOrbitCoupled2017,liviSyntheticDimensionsSpinOrbit2016,chalopinProbingChiralEdge2020}, Rydberg states~\cite{kanungoRealizingTopologicalEdge2022,luProbingTopologicalPhase2024,luWavepacketDynamicsLongrange2024,trautmannRealizationTopologicalThouless2024,chenQuantumWalksCorrelated2024,chenStronglyInteractingRydberg2024}, quantised momentum states~\cite{lohseExploring4DQuantum2018,viebahnMatterWaveDiffractionQuasicrystalline2019,meierObservationTopologicalSoliton2016,meierObservationTopologicalAnderson2018} and atomic trap states~\cite{oliverBlochOscillationsSynthetic2023}. However, the rotational states of polar molecules offer a unique set of advantages as they are connected via strong electric-dipole transitions in microwave domain and  benefit from dipolar interactions between molecules, while being immune to radiative decay losses. This opens up the possibility of preparing exotic many-body ground states using molecular synthetic dimensions~\cite{sundarStringsUltracoldMolecules2019,sundarSyntheticDimensionsUltracold2018}.

\begin{figure}[t!]
\centering
\includegraphics[width=\linewidth]{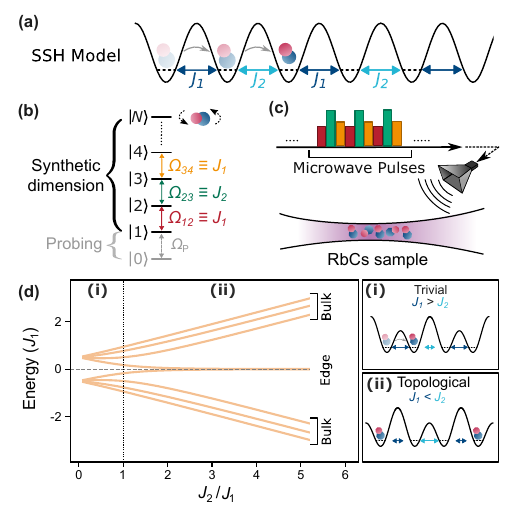}
\caption{\textbf{Encoding the SSH model in the rotational structure of RbCs molecules.}{ (a) An illustration of a lattice with the SSH model showing the alternating tunnelling rates $J_1$ and $J_2$. (b) The SSH Hamiltonian is encoded onto a synthetic lattice corresponding to the $\ket1$ to $\ket{N}$ rotational states (where $N$ is the rotational quantum number). The sites are coupled using microwave fields with Rabi frequencies $\Omega_{NN'}$. The ground state $\ket{0}$ is used as an auxiliary probe state. (c) The target Hamiltonian is implemented stroboscopically by interleaving resonant microwave pulses which are applied to the sample of molecules confined in an optical trap. (d) The energies of the eigenstates of the 8-level SSH Hamiltonian as a function of $J_2/J_1$. The SSH model exhibits a topological phase transition in the thermodynamic limit at $J_2/J_1 = 1$, from (i) a trivial to (ii) a topological regime. This is characterised by the appearance of the two edge eigenstates.}} 

\label{fig:SSH_Scheme}
\end{figure}

Perhaps the simplest synthetic dimension is a 1D tight-binding model, which can be encoded in a chain of microwave-coupled molecular rotational states. Such a chain can be used to implement the well-known Su-Schrieffer-Heeger (SSH) model that was originally developed to understand the electrical conductivity of trans-polyacetylene~\cite{suSolitonsPolyacetylene1979}. The model is characterised by a 1D chain comprising an even number of sites coupled with alternating strengths $J_1, J_2$ as shown in Fig.\,\ref{fig:SSH_Scheme} (a). Despite its simplicity, the SSH model is known to exhibit non-trivial topological features. Moreover, it has served as an excellent benchmark of various experimental platforms, having been studied in both real-space lattices~\cite{deleseleucObservationSymmetryprotectedTopological2019,atalaDirectMeasurementZak2013} and synthetic dimensions~\cite{kanungoRealizingTopologicalEdge2022,meierObservationTopologicalSoliton2016,liDirectExtractionTopological2023,wangVersatilePhotonicFrequency2025}.


Here, we report on the realisation of a 1D chain synthetic dimension within the rotational states of ultracold $^{87}$Rb$^{133}$Cs molecules (hereafter RbCs) and study SSH models with up to 8 sites. We probe the phase transition between trivial and non-trivial topological phases by tuning the tunnelling rates in the synthetic lattice. We experimentally measure the energies of the SSH eigenstates using a combination of spectroscopic and interferometric methods. We demonstrate the topological protection of the SSH edge states against chiral perturbations, and finally extract the winding number from the dynamics of the system, showing that it changes across the topological phase transition. Our results show excellent agreement with theoretical predictions from exact diagonalisation over many tunnelling periods, highlighting our ability to engineer precisely controlled tunnelling rates between lattice sites and the long coherence times available in this system that enable the observation of coherent dynamics.


\begin{figure*}[t!]
\centering
\includegraphics[]{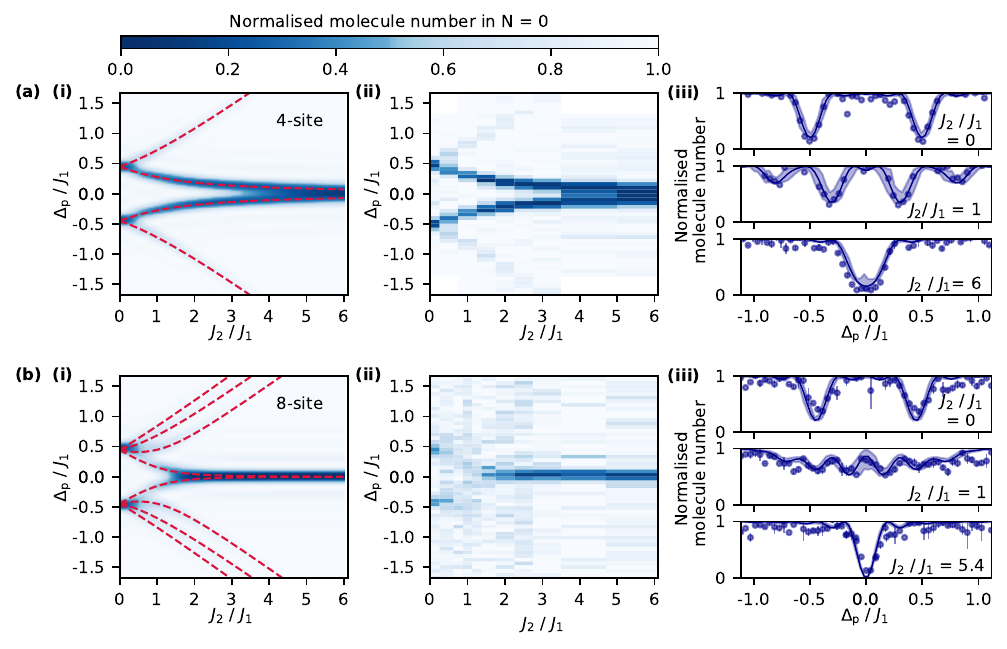}
\caption{\textbf{Spectroscopy of the dressed eigenstates of the SSH Hamiltonian.} We observe loss of molecules from $N=0$ when the probe field is tuned close to resonance to an eigenstate. Results are shown for (a) 4-site and (b) 8-site SSH chains. For each, we show (i) a theoretical prediction and (ii) experimental measurements of the normalised number of molecules remaining in $N=0$ as a function of probe detuning and the SSH tunnelling ratio $J_2/J_1$. In panel (i), the red dashed lines indicate the eigenenergies calculated by exact diagonalisation of the applied Hamiltonian. Panel (iii) shows the experimental measurements at specific ratios of $J_2/J_1$ (indicated inset) corresponding to the system in the trivial phase, at the transition boundary, and in the non-trivial topological phase. The solid lines indicate the theoretical predictions at the calibrated Rabi frequencies, with the shaded regions denoting the 1-$\sigma$ error surface. Error bars on all plots show 1$\sigma$ confidence intervals.}
\label{fig:SSH_spectroscopy}
\end{figure*}


\subsection*{Realising a 1D synthetic chain in RbCs}


We encode a 1D synthetic dimension into the rotational states of RbCs molecules, with the first rotationally excited state ($N=1$) as the first site in the 1D chain, as shown in Fig.\,\ref{fig:SSH_Scheme}(b). We label the rotational states (and their corresponding synthetic sites) by the angular momentum quantum number $N$. We use the spin-stretched states where the rotational angular momentum projection and both nuclear spin projections take their maximum values, i.e. $m_N=N$, $m_\mathrm{Rb}=I_\mathrm{Rb}=3/2$, $m_\mathrm{Cs}=I_\mathrm{Cs}=7/2$ where $m_\mathrm{Rb}, m_\mathrm{Cs}$ are the projections of the nuclear spins $I_\mathrm{Rb},I_\mathrm{Cs}$.


To drive multiple rotational transitions in a scalable way, we employ a stroboscopic approach, rapidly alternating between driving each transition in turn using a single microwave source, as illustrated in Fig.~\ref{fig:SSH_Scheme}(c). The target Hamiltonian is realised as the time average of many pulses, each of which is 40 to 200 times shorter than the characteristic tunnelling times (see Methods). By controlling the amplitudes of the microwave fields we control the effective tunnelling rates, $J_1$ and $J_2$, in the synthetic dimension, shown in Fig.~\ref{fig:SSH_Scheme}(a). This approach is an example of Floquet engineering~\cite{choiRobustDynamicHamiltonian2020,geierFloquetHamiltonianEngineering2021} or equivalently
Trotterisation, as would be required in a gate-based simulator~\cite{lloydUniversalQuantumSimulators1996}. As discussed in the Methods, the Trotter infidelity between the target and applied Hamiltonians is negligible, of the order $10^{-4}$ for the pulse durations used in this work.



Probing the synthetic chain with high spectroscopic resolution requires long interrogation times and long-lived coherence between rotational states. We therefore confine the molecules in a near-magic-wavelength optical trap~\cite{ gregorySecondscaleRotationalCoherence2024,ruttleyLonglivedEntanglementMolecules2025}, where tuning the wavelength of the trap allows control of the differential light shifts between rotational states in order to maximise the coherence time.~\cite{guanMagicConditionsMultiple2021,hepworthLonglivedMultilevelCoherences2025}. The exact wavelength used for the optical trapping is tailored to each experiment,  as described in the Methods. We note that although we can not reach the exact magic condition for all states used, we conservatively estimate coherence times across 8 rotational states that are two orders of magnitude larger than the tunnelling times in the synthetic lattice.  

\begin{figure*}[t!]
\includegraphics{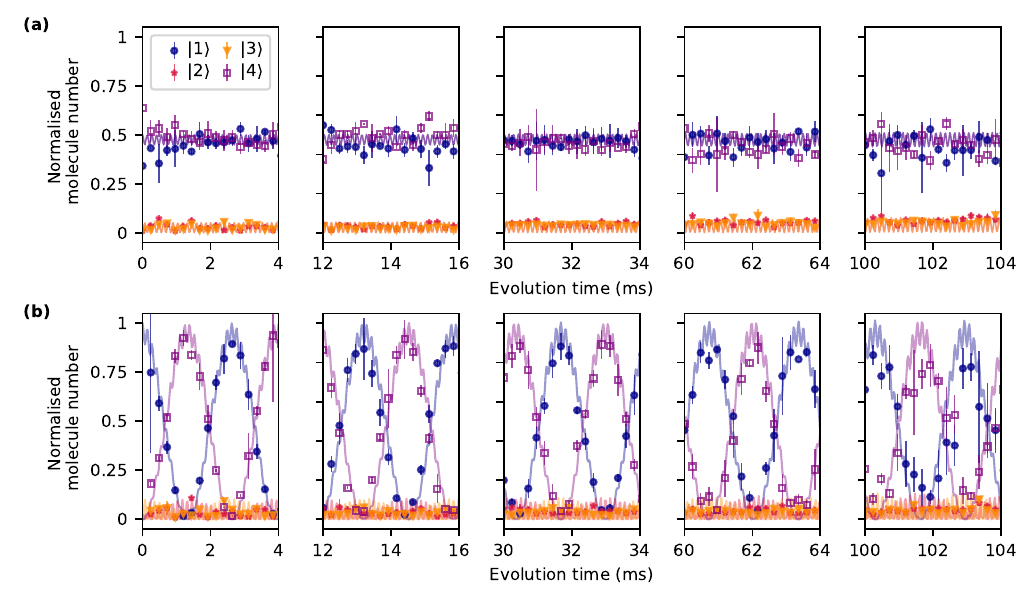}
\caption{\textbf{The time evolution of the site populations under the 4-site SSH Hamiltonian after initialising in (a) an edge eigenstate and (b) a superposition of the edge eigenstates.} (a) We initialise the molecule in the state $\ket{\Phi_{e+}} = \frac{1}{\sqrt{2}} (\ket{1}$ + $\ket{4})$ and evolve under the SSH Hamiltonian with $J_1= 2.30(2)$\,kHz and $J_2 = 13.9(2)$\,kHz. We see that the population remains equally split between $\ket1$ and $\ket4$, indicating that $\ket{\Phi_{e+}}$ is a good approximation of an eigenstate. (b) Keeping $J_1$ and $J_2$ the same, we initialise in site $\ket{1}$, which is a linear superposition of the two edge eigenstates $ \frac{1}{\sqrt{2}} (\ket{\Phi_{e+}}  + \ket{\Phi_{e-}}$), and again evolve under the SSH Hamiltonian. The population now oscillates between $\ket1$ and $\ket4$ with a frequency determined by $E_{\Phi_{e+}} - E_{\Phi_{e-}}$. From a fit to the data, we obtain an oscillation frequency of 379.0(7)\,Hz, in agreement with the predicted value of 371(13)\,Hz. The solid lines in both (a) and (b) are the theoretical predictions for $J_1 = 2.31$\,kHz and $J_2 = 13.7$\,kHz. Error bars on all plots show 1$\sigma$ confidence intervals. } 
\label{fig:SSH_time}
\end{figure*}

\subsection*{Probing the edge states of the SSH Hamiltonian}

We use our molecular synthetic dimension to probe the topological phase transition of the SSH model using two complementary approaches: spectroscopic measurements of the eigenspectrum and observation of the time dynamics. To set the context for both, we first consider the eigenenergies of the SSH Hamiltonian for a finite chain, shown in
Fig.\,\ref{fig:SSH_Scheme}(d), where the energies are plotted as a function of the ratio of the tunnelling amplitudes $J_2/J_1$. The topological phase transition occurs at $J_2/J_1 = 1$. When $J_2/J_1 < 1$, the system is in the trivial (or normal) phase. When $J_2/J_1 > 1$, the system enters a non-trivial topological phase, characterised by the emergence of two edge eigenstates which are localised on the ends of the chain. In an $n$-site SSH chain, these edge eigenstates are given by the hybridisation of the odd $\left(\ket{\phi_{\text{odd}}}=\sum_{i=1}^{n/2} a_{2i-1}\ket{2i-1}\right)$ and even $\left(\ket{\phi_{\text{even}}}=\sum_{i=1}^{n/2} a_{2i}\ket{2i}\right)$ modes, where $a_{2i-1}/a_1 = (-1)^{i-1}e^{-(i-1)/\xi}$ and $a_{2i}/a_n = (-1)^{n/2-i}e^{-(n/2-i)/\xi}$ with $\xi = 1/\ln(J_2/J_1)$ being the edge localisation length. 

To realise an SSH chain, we set the ratios of tunnelling rates to have alternating strengths. We begin with a 4-site chain, the shortest chain compatible with the SSH geometry, by setting the Rabi frequencies $\Omega_{NN'}$ coupling neighbouring states~$N, N'$ such that $\Omega_{12}=J_1$, $\Omega_{23}=J_2$ and $\Omega_{34}=J_1$. We investigate the energy level structure for a range of tunnelling rate ratios, fixing $J_1 = 2.3(2)$\,kHz and varying $J_2$, by performing spectroscopy using the auxiliary $N=0$ probe level as shown in Fig.\,\ref{fig:SSH_Scheme}(b). With the molecules prepared in $N=0$, we apply a weak probe field with a bare Rabi frequency $\Omega_\mathrm{p}=257(12)\,\mathrm{Hz}$ that addresses site $\ket1$. The probe field is also pulsed stroboscopically, alternating with the SSH driving fields. We set the total probe time to be approximately  $\pi/\Omega_\mathrm{p}$. 

We measure the number of molecules remaining in $N=0$ as a function of the detuning of the probe field $\Delta_{\text{p}}$ for a range of $J_2/J_1$. The results are shown in Fig.\,\ref{fig:SSH_spectroscopy}(a). With the SSH driving fields present in the sequence, the molecule number drops when the probe is resonant with a transition to one of the dressed eigenstates of the SSH Hamiltonian. The transition strength to each dressed state is reduced from that of the bare transition, and is given by Fermi's golden rule such that the on-resonance Rabi coupling to a given dressed state $\beta$ is $\Omega'_\mathrm{p} = \Omega_\mathrm{p}\braket{\beta}{1}$. The spectral width of the coupling originates from the finite probe pulse time. In the limit where the number of sites $n\rightarrow\infty$, this measurement is analogous to measuring the local density of states at site~$\ket{1}$ of the synthetic chain.

Although the 4-site chain is minimal, we observe the key features associated with the topological phase transition of the SSH model. When $J_2/J_1=0$, we see two dips in the populations at $\Delta_\text{p} = \pm J_1/2$ corresponding to even and odd superpositions of the neighbouring sites $(\ket{1} \pm \ket{2})/\sqrt{2}$. This trivial case is identical to a 3-level Autler-Townes system as previously observed with RbCs rotational transitions in~\cite{blackmoreCoherentManipulationInternal2020}. At the transition boundary, \(J_2/J_1=1\), all the SSH eigenstates are hybridised across all the lattice sites, and we see 4 distinct peaks in the central panel of Fig.~\ref{fig:SSH_spectroscopy}(a)(iii). As $J_2/J_1$ is increased, we see the localisation of two of the eigenstates onto $\ket{1}$. These eigenstates correspond to the SSH edge eigenstates. When $J_2\gg J_1$, we can approximate the edge eigenstates as the even and odd superpositions of the sites at the two ends of the chain, $ \ket{\Phi_\mathrm{e\pm}} = (\ket1 \pm \ket4)/\sqrt{2}$. As expected, the energies of these eigenstates tend towards $\Delta_\mathrm{p}=0$ as $J_2/J_1 \rightarrow \infty$.


Increasing the length of the chain leads to a more pronounced phase transition and smaller energy-splitting between the edge states. We therefore increase the chain length to 8 sites, by coupling rotational states up to $N=8$. The results are shown in Fig.~\ref{fig:SSH_spectroscopy}(b). For the 8-site chain, we use a lower tunnelling rate $J_1 = 0.47(2)$\,kHz and probe Rabi frequency $\Omega_\mathrm{p}= 62(4)$\,Hz. Now at the phase transition $J_2/J_1=1$, the wavefunctions of all 8 eigenstates have similar contributions from site $\ket1$. This leads to each eigenstate having little transition strength and results in small variations in the number of detected molecules over a broad range of probe detunings, as shown in the central panel of Fig.~\ref{fig:SSH_spectroscopy}(b)(iii). As $J_2/J_1$ is increased, the two edge eigenstates strongly localise onto site $\ket1$, and quickly become too close together in energy to resolve using our probe parameters. With $J_2/J_1 = 5.4$, this energy splitting is predicted to be $3.1\,\text{Hz}\simeq J_1/150$, whereas the Fourier-limited resolution of the probe pulse is $124(8)\,\text{Hz}\simeq J_1/3$ . Later we show how such small energy splittings can be measured using an interferometric technique.

Next we study the time dynamics of the molecules under the 4-site SSH Hamiltonian. To do this, we prepare the molecules in a chosen initial state using a series of coherent $\pi$ and $\pi/2$ pulses~\cite{gregoryControllingRotationalHyperfine2016}. We then switch on the SSH driving fields in the topological regime with $J_2/J_1\approx6$ and allow the system to evolve for a variable time, after which we measure the number of molecules present in each rotational state (see Methods). The results are shown in Fig.~\ref{fig:SSH_time}. 

We begin by preparing the molecules in the superposition $\ket{\Phi_\mathrm{e+}}=(\ket1+\ket4)/\sqrt{2}$. This is identical to one of the edge states in the limit $J_2/J_1\rightarrow\infty$. The time evolution for this initial state is shown in Fig.~\ref{fig:SSH_time}(a). We see that the populations in sites $\ket1$ and $\ket4$ remain invariant with time to within the experimental noise, indicating that $\ket{\Phi_{e+}}$ is a good approximation of an eigenstate. Indeed, we calculate the overlap between the prepared state $\ket{\Phi_\mathrm{e+}}$ and the actual eigenstate of the system with $J_2/J_1=6$ to be 0.973.

Next, we prepare the population in site~$\ket{1}$. This is not an eigenstate of the SSH Hamiltonian and is projected primarily onto an equal superposition of the two edge eigenstates. We observe that the population undergoes long-range tunnelling between sites $\ket{1}$ and $\ket{4}$, as shown in Fig.\,\ref{fig:SSH_time}(b), as the relative phase of the two edge eigenstates evolves with time. The frequency of these oscillations is equal to $\Delta_\text{e}$, the energy splitting between the two edge states. This interferometric technique allows us to measure the energy splitting $\Delta_\text{e}=379.0(7)$\,Hz with Hertz-level precision.

The long coherence times afforded by our magic-wavelength trap allow measurements out to evolution times greater than 100\,ms. Crucially, the coherence times we achieve are significantly larger than the tunnelling times of the driven transitions $\tau_\text{1} = 2/J_1 \simeq 213$\,\unit{\micro\second} and $\tau_2 = 2/J_2 \simeq 36$\,\unit{\micro\second}. Previous measurements of long-range tunnelling in synthetic dimensions realised with Rydberg atoms were limited to coherence times $\sim10\,\tau_1$ \cite{luWavepacketDynamicsLongrange2024}. Here, we see minimal dephasing over significantly longer timescales, up to $\sim 500 \,\tau_1$. This is an advantage of molecules that allows precise measurements of Hamiltonians engineered in synthetic dimensions.

\begin{figure}[t!]
\centering
\includegraphics{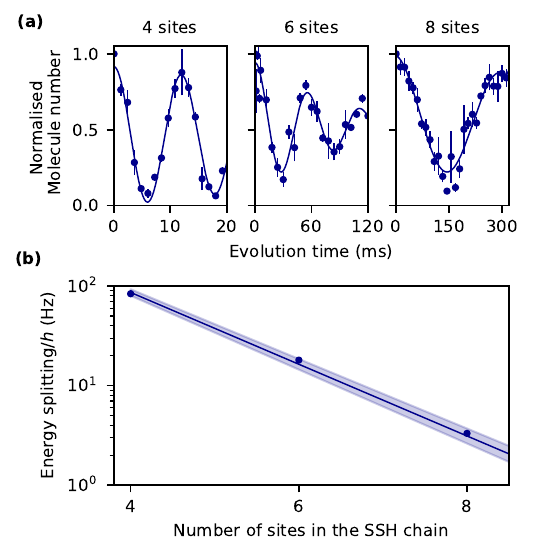}
\caption{\textbf{Dependence of the energy splitting between the edge states on the length of the SSH chain.}{ (a) We study dynamics in SSH chains of varying length, with $J_1 = 470(15)$\,Hz and $J_2 = 2.5(1)$\,kHz, and the system initialised in site $\ket1$. In each case, the evolution of the population in site $\ket1$ with time is plotted and the energy splitting of the edge states is extracted by fitting the observed oscillations to a damped sinusoid. (b)  The fitted edge-state energy splitting as a function of the number of sites in the SSH chain. 
As expected, we observe that the energy splitting reduces exponentially as the length of the SSH chain increases, with the blue line showing the expected behaviour, and the shaded area the 1-$\sigma$ error bounds from the uncertainties in Rabi frequencies. Error bars on all plots show 1$\sigma$ confidence intervals.}}
\label{fig:SSH_N_levels}
\end{figure}

\begin{figure*}[t!]
\includegraphics{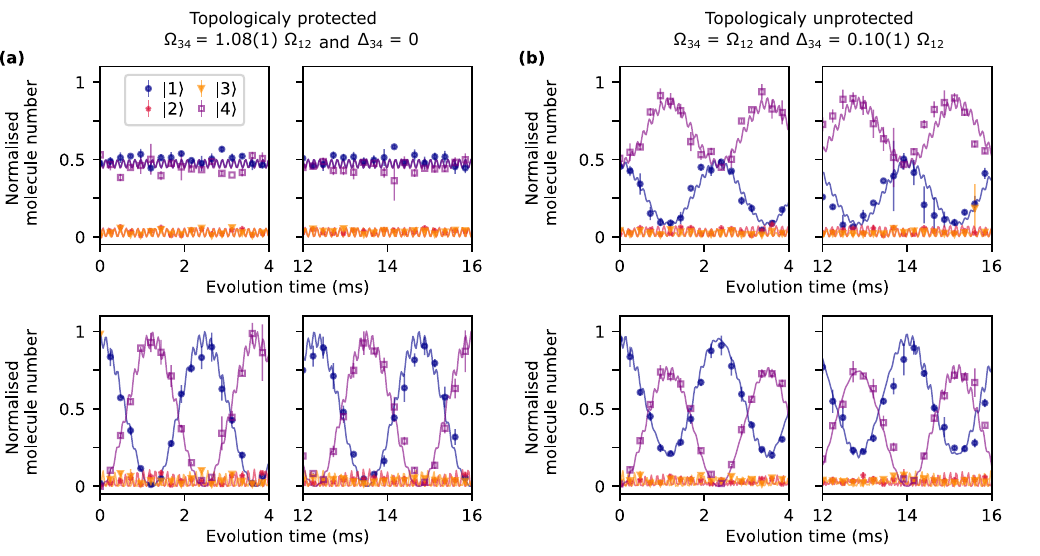}
\caption{\textbf{Topological protection of the edge eigenstates in the 4-site SSH Hamiltonian.} We implement perturbed versions of the 4-site SSH Hamiltonian, and show that the edge eigenstates are protected against chiral perturbations, but not against non-chiral ones. We again initialise the molecules in the SSH eigenstate $\ket{\Phi_\text{e+}}$ (upper panels) and site $\ket{1}$ (lower panels), and plot the population dynamics for all four sites. The unperturbed tunnelling rates $J_1$ and $J_2$ are the same as in Fig. \ref{fig:SSH_time}. (a) A chiral perturbation, with imbalanced Rabi frequencies. We see that the state $\ket{\Phi_\text{e+}}$ remains a good approximation of the eigenstate, as evidenced by the absence of population dynamics in the upper panels. Comparing the frequency of the oscillations in the lower panels with that from Fig.\,\ref{fig:SSH_time} shows that the eigenergies are altered by the perturbation. (b) A non-chiral perturbation, with $\Delta_\text{34} \neq 0$, which raises the energy of site $\ket{4}$. The significant oscillations when the system is initialised in $\ket{\Phi_\text{e+}}$ show that $\ket{\Phi_\text{e+}}$ is no longer a good representation of the eigenstate. The solid lines in both (a) and (b) show the theoretical predictions for the evolution of the populations at the experimental parameters. Error bars on all plots show 1$\sigma$ confidence intervals.}

\label{fig:SSH_topology}
\end{figure*}

 For an $n$-site SSH chain, the energy splitting between the edge states is predicted to decrease exponentially with $n$, such that $\Delta_\text{e}\sim (J_1/J_2)^{n/2}$. We test this by projecting the initial state $\ket{1}$ onto SSH chains of varying length, with $J_1 = 470 (15)$\,Hz and $J_2 = 2.5(1)$\,kHz. In each case, we measure the population in site $\ket1$ as a function of time, as shown in Fig.\,\ref{fig:SSH_N_levels}(a), and fit the oscillations using a damped sine wave to extract $\Delta_\text{e}$. For consistency, we keep the wavelength of the magic trap fixed at the value found to be optimum for the 8-site chain, resulting in coherence times for the 4-site and 6-site cases that are below their individually optimised values. We find that as the lattice length increases, the energy splitting decreases exponentially, as shown in Fig.\,\ref{fig:SSH_N_levels}(b). The smallest energy splitting we observe is 3.31(13)\,Hz for the 8-site chain, and the dependence on the chain length is in good agreement with the theoretical expectation. 
 
 
 

\subsection*{Topological protection of the edge states}

The edge states of the SSH Hamiltonian are topologically protected from perturbations that preserve chiral symmetry, while remaining susceptible to perturbations that do not. The unperturbed SSH model is bipartite; odd and even sites form sublattices, that we label $A$ and $B$, respectively, and the coupling is only between the two sublattices. In this basis, the Hamiltonian is of the form $\mathcal{H}_{\text{SSH}} = \mathcal{M}_{AB}\ket{A}\bra{B} + \mathcal{M}^\dagger_{AB}\ket{B}\bra{A}$, where $\ket{A} = \sum_{i=1}^{n/2}\ket{2i-1}$ and $\ket{B} = \sum_{i=1}^{n/2}\ket{2i}$, and $\mathcal{M}_{AB}$ is the matrix element in the sublattice basis ${A,B}$. The Hamiltonian obeys the anti-commutation relation $\{\mathcal{H}_{\text{SSH}},\Gamma\}=0$, where $\Gamma = \ket{A}\bra{A} - \ket{B}\bra{B}$ is the chiral operator. Perturbations of the Rabi frequencies only modify the $\mathcal{M}_{AB}$ terms without adding additional terms in the bipartite basis. In the thermodynamic limit, these perturbations leave the twofold degeneracy of the zero-energy eigenstates unchanged\,\footnote{For a finite system, there are two approximate zero-energy states with an energy gap that is exponentially small in the system size. Under chiral perturbations, there still exist two states which are exponentially close to zero energy.}. However, adding non-uniform detunings introduces terms like $\ket{A}\bra{A}$ and $\ket{B}\bra{B}$, in which case the chiral symmetry no longer holds, and the edge states are no longer present. We experimentally examine the effects of chiral (perturbed Rabi frequencies) and non-chiral (non-uniform detunings) perturbations by repeating the measurements in Fig.\,\ref{fig:SSH_time} whilst applying perturbed versions of the 4-site SSH Hamiltonian. 

To realise a chiral perturbation, we break the symmetry of the tunnelling rates by setting $\Omega_{12} \neq \Omega_{34}$. Fig.\,\ref{fig:SSH_topology}(a) shows the time dynamics of molecules initialised in $\ket{\Phi_\text{e+}}$ (upper panels) and $\ket{1}$ (lower panels) with a coupling ratio of $\Omega_{34}/\Omega_{12}=1.08(1)$. We see that, as before, when the population is initialised in the state $\ket{\Phi_\text{e+}}$, the state composition remains unaffected by the perturbed SSH Hamiltonian. This indicates that the state $\ket{\Phi_\text{e+}}$ remains a good approximation of an eigenstate of the system. When the system is initialised in site $\ket{1}$, and evolved under this perturbed Hamiltonian, we see oscillations as before, but with an increased frequency of 406(3)\,Hz compared to the unperturbed SSH Hamiltonian. This increase in energy splitting agrees well with our theoretical simulations shown by the solid lines in Fig. \ref{fig:SSH_topology}. We use these simulations to estimate the fidelity ($\mathcal{F}$) between the perturbed and unperturbed eigenstates, finding $1- \mathcal{F} \approx 2 \times 10^{-4}$.

For a non-chiral perturbation, the microwave driving the $\ket2 \rightarrow \ket3$ transition is detuned from resonance to effectively add a local potential offset to site~$\ket{4}$ of the synthetic chain, thereby introducing a non-zero diagonal term to the Hamiltonian. Figure \ref{fig:SSH_topology} (b) shows the time dynamics of molecules initialised in $\ket{\Phi_\text{e+}}$ (upper panels) and $\ket{1}$ (lower panels) when evolved under a non-chiral perturbation with $\Delta_\text{34} = 0.10(1)\,J_1$.  Now, when we initialise in $\ket{\Phi_\text{e+}}$,  we see large population oscillations, indicating that $\ket{\Phi_\text{e+}}$ is no longer a good approximation of an eigenstate of the system. When the system is initialised in site $\ket1$, the population oscillations no longer have full contrast, as $\ket1$ is no longer an equal superposition of two eigenstates of the system. The fidelity between the perturbed and unperturbed edge eigenstates is calculated again, with the error $1- \mathcal{F} = 0.021(2)$, which is significantly larger than the chirally perturbed case.

\begin{figure}[t!]
\centering
\includegraphics{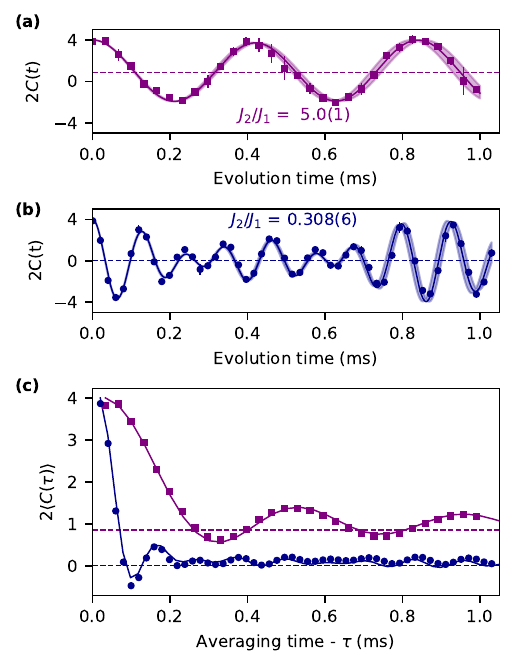}
\caption{\textbf{Measurement of the mean chiral displacement in a 4-site SSH chain.} 
(a) Initialising in site $\ket{3}$, we measure twice the chiral displacement, $2C(t)$, in the topological case with $J_2/J_1 =5.0(1)$ using complete site-resolved detection. (b) The same measurement in the trivial case with $J_2/J_1=0.308(6)$. In (c), we plot twice the mean chiral displacement, $2\langle{C(\tau)}\rangle$, obtained by averaging $C(t)$ in (a) and (b) over an increasing time window, $\tau$. For the infinite-site SSH model, $2\langle{C(\tau)}\rangle_{\tau\rightarrow\infty}$ converges to the winding number: $\mathcal{W}=1$ in the topological regime and $\mathcal{W}=0$ in the trivial regime. For the 4-site system studied here, we expect $2\langle{C (\tau)}\rangle_{\tau\rightarrow\infty}$ to be 0.02 for the trivial case and 0.86 for topological case. These values are indicated by the dashed horizontal lines in each panel. The solid lines are the theory predictions for the applied SSH parameters with shaded areas denoting 1$\sigma$ confidence intervals. Error bars on all plots show 1$\sigma$ confidence intervals. }
\label{fig:SSH_winding}
\end{figure}

\subsection*{Measurement of the winding number}

The winding number, $\mathcal{W}$, of the Bloch states around the Brillouin zone is a quantised topological invariant that characterises one-dimensional chiral lattice models. For a chiral symmetric model, the number of edge states is given by $2\abs{\mathcal{W}}$~\cite{asboth_short_2016}. For the SSH model in the thermodynamic limit, we expect that $\mathcal{W} =0$ in the trivial regime and $\mathcal{W} = 1$ in the topological regime. 

The winding number of the system can be extracted experimentally by finding the mean chiral displacement, $\langle C(\tau) \rangle$, in the long time limit, as
$\mathcal{W} \approx 2\langle C(\tau)\rangle_{\tau\rightarrow\infty}$~\cite{maffeiTopologicalCharacterizationChiral2018}. The chiral displacement of the system at time $t$ is defined as $C(t) = \Gamma\hat{m}(t)$, where $\Gamma$ is the chiral operator defined earlier and $\hat{m}$ is the position operator that labels the position of a lattice site in unit cells, as defined in~\cite{maffeiTopologicalCharacterizationChiral2018}. Each unit cell of the SSH model consists of adjacent sites (one odd and one even), resulting in a diagonal matrix $\hat{m}=\mathrm{diag}(1,1,2,2...)$. Evaluating $C(t)$ for the 4-site SSH thus yields
\begin{equation}\label{eqn: chiral disp}
    C(t) = P_{\ket1} (t) - P_{\ket2} (t) +2[P_{\ket3}(t) - P_{\ket4} (t)]
\end{equation}
where $P_{\ket{N}}(t)$ is the fractional population in site $\ket{N}$.

We measure the chiral displacement for the 4-site SSH chain in the trivial and topological regimes. We initialise the system in site~$\ket3$ and measure the population of each site as a function of the evolution time under the SSH driving fields. We then use Eq.\,\ref{eqn: chiral disp} to extract the chiral displacement as a function of time. Here, we use tunnelling ratios of $J_2/J_1=0.308(6)$ for the trivial regime and $J_2/J_1=5.0(1)$ for the topological regime. Our results are shown in Fig.\,\ref{fig:SSH_winding}(a,b), alongside a no-free-parameter model, shown by the solid lines. We see periodic oscillations in the chiral displacement for both the trivial and topological regimes, but in the topological case (Fig.\,\ref{fig:SSH_winding}(a)) the oscillations are offset from zero.

In Fig.\,\ref{fig:SSH_winding}(c), we show the mean chiral displacement as a function of time, obtained by averaging the measured chiral displacements shown in Fig.\,\ref{fig:SSH_winding}(a) and (b) over an increasing time window, i.e. $\langle C(\tau)\rangle = 1/\tau \int_0^\tau C(t)dt$. The measurements are performed over a time of 1\,ms, larger than the characteristic oscillation periods in both cases. The solid lines in Fig.\,\ref{fig:SSH_winding} again show the no-free-parameter model, accounting for the discrete times at which the experimental measurements are performed.  The root-mean-square errors (RMSE) between the theory and experiment are 0.06 for the topological case and 0.09 for the trivial case, where RMSE is defined as $\sqrt{\frac{1}{n}\sum_{i=1}^n(x_i -x'_i)^2}$, confirming close agreement with the expected behaviour. In each case, the mean chiral displacement converges to a distinct value. In the topological case, twice the mean chiral displacement is expected to converge to 0.86, while in the trivial case it converges to 0.02. These values are shown in Fig.\,\ref{fig:SSH_winding} by the dashed lines. The discrepancy from $\mathcal{W} = 0~\text{or}~1$ is due to the finite size of the system.

\section{Outlook}

The experiments we present are the first realisation of a synthetic dimension in ultracold polar molecules. While the 1D chain geometry we encode here is relatively simple, more complex and arbitrary patterns of coupling strength and detuning along the chain can be easily achieved \cite{ozawaTopologicalQuantumMatter2019}. Moreover, utilising the hyperfine structure of the molecules would allow more complex synthetic lattices, including 2D lattices and loop geometries simulating 1D chains with periodic boundary conditions~\cite{sundarSyntheticDimensionsUltracold2018}. The long coherence times available could also enable adiabatic preparation of ground states~\cite{born1928beweis,PhysRevB.91.134303,venegas2020adiabatic} in many-body molecule systems with synthetic dimensions. In related theoretical work, we have developed computational techniques for identifying ideal sets of states for different synthetic lattice geometries~\cite{hepworthIsolatedQuantumstateNetworks2025}. 

Interesting possibilities arise from combining real and synthetic dimensions in ensembles of molecules. Real-space dipolar interactions allow the exchange of rotational energy between molecules, with interaction strengths that depend on the relative positions of the molecules in the ensemble. Sundar~\emph{et~al.} have studied the combination of 1D and 2D real-space dimensions with 1D synthetic chains.~\cite{sundarStringsUltracoldMolecules2019} For molecules pinned to sites in the real dimension, they predict the spontaneous formation of 1D synthetic strings or membranes when the dipolar interactions are strong compared to the couplings along the synthetic chain. Beyond this, ultracold molecules uniquely allow the combination of dipolar interactions with tunnelling between sites in a real-space lattice; the work we have presented demonstrates that coherence in the synthetic dimension can persist for sufficiently long times that molecular motion could affect the dynamics.

In conclusion, we have encoded a coupled 1D chain of lattice sites in the rotational states of ultracold RbCs molecules. We have used this to study the SSH model for up to 8 sites, demonstrating and benchmarking the use of molecules for engineering synthetic dimensions. We have observed the phase transition between trivial and non-trivial phases, characterised by a change in the winding number and the emergence of topologically protected edge states. Our experiments pave the way to a new paradigm for quantum simulation with ultracold molecules where the internal degrees of freedom of the molecule provide the means to study new models with complex geometries, synthetic gauge fields or mixed real and synthetic dynamics.

\section{Methods}

\subsection*{Preparation and state-sensitive detection of ultracold RbCs molecules}

We prepare ultracold RbCs molecules in a single hyperfine level of the $\mathrm{X}^1\Sigma$ ground state by association from a pre-cooled mixture of Rb and Cs atoms; details of the molecule preparation and imaging can be found in~\cite{mortlockMultistateDetectionSpatial2025,matthiesLongdistanceOpticalconveyorbeltTransport2024}. In brief, we merge optically trapped samples of ultracold Rb and Cs atoms to form a high phase-space-density mixture. We then create molecules by magnetoassociation via a magnetic field ramp across an interspecies Feshbach resonance at 197\,G~\cite{molonyProductionUltracold872016}. These weakly bound molecules are then transferred optically to their rovibronic ground state using stimulated Raman adiabatic passage~\cite{vitanovStimulatedRamanAdiabatic2017,bergmannCoherentPopulationTransfer1998} and the remaining atoms removed using resonant laser light. The ground-state transfer is performed in free space by briefly switching off the trap light to avoid spatially varying light shifts that otherwise limit the efficiency of this process. Following the ground-state transfer, the molecules are recaptured in the same optical trap as the initial atomic mixture.   

We detect the molecules using the site-resolved fluorescence imaging method we reported in~\cite{mortlockMultistateDetectionSpatial2025}.
After the microwave interrogation, the molecules are pinned in a deep 1064\,nm optical lattice (trap depths of $\approx 400$\,\unit{\micro\kelvin} for RbCs), and then dissociated to form atom pairs by reversing the STIRAP and sweeping the magnetic field back across the 197\,G Feshbach resonance. After removal of the Rb atoms with resonant light, we image the remaining Cs atoms by detecting fluorescence from $D_2$ optical molasses. A high-numerical-aperture objective allows us to detect the locations of individual Cs atoms in the pinning lattice, and each atom marks the lattice site previously occupied by a RbCs molecule. We typically detect around 400 molecules, and observe a background signal of $3^\text{+5}_\text{-2}$ Cs atoms in the absence of the return STIRAP pulse. We ascribe this background predominantly to the imperfect removal of the Cs atoms used to create the molecules. 

Crucially, STIRAP is state selective, therefore only molecules in the $\ket{0}$ state are dissociated to an atom pair and thus detected. To detect molecules in other rotational states we must transfer them to $\ket{0}$ with a series of coherent $\pi$-pulses before they are pinned. We measure $\pi$-pulse fidelities of $\geq$~99\% for the pulses used to initialise and read out all the different molecular states by comparing the molecule numbers in sequences with and without an additional 10 $\pi$-pulses per transition. We typically utilise Rabi frequencies of $\approx10$\,kHz for the initialisation and readout pulses.

\subsection*{Magic-wavelength optical dipole trap}

The molecules are confined to an optical dipole trap formed by a single beam. This beam is linearly polarised parallel to the applied magnetic field of 181.6\,G, and has an elliptical profile characterised by beam waists of 8\,\unit{\micro \metre} vertically and 150\,\unit{\micro\metre} horizontally. This geometry is chosen to reduce the intensity required to hold the molecules against gravity. The molecules are held in this trap for 50~ms after association to allow them to equilibrate with the trap, while all other trapping and cooling beams are blocked by mechanical shutters. The molecules then undergo microwave interrogation according to the experimental sequence. The lifetime of the molecules in this trap is $>1$\,s.

The trap light is frequency-stabilised with reference to a high-finesse optical cavity with an ultra-low expansion glass spacer. The frequency of the light is $\approx185$\,GHz blue-detuned from the ($\mathrm{X}^1\Sigma^+,v=0,N=1,M_\text{F} = 6$) to ($\mathrm{b}^3\Pi_0,v=0,N'=1,M_\text{F}' = 5$) transition. By carefully tuning the laser frequency we can eliminate the differential polarisability between a given pair of rotational states~\cite{guanMagicConditionsMultiple2021,gregorySecondscaleRotationalCoherence2024,ruttleyLonglivedEntanglementMolecules2025}. However, the trap cannot be made exactly magic for all rotational transitions at the same time~\cite{hepworthLonglivedMultilevelCoherences2025}. 

For the spectroscopy measurements in Fig.\,\ref{fig:SSH_spectroscopy}, we choose the laser frequency to maximise the shortest coherence time present in the chain. A different laser frequency is required to achieve this for the 4-site chain compared to the 8-site chain. In each case, the maximum differential light shift is present between the states at the ends of the chain. For the 4-site chain, we use a detuning of 185.243~GHz. Assuming decoherence solely from thermal motion in the trap, with temperature $T\approx 2$\,\unit{\micro\kelvin} and peak trap intensity $I \approx 6.5\,\text{kW/cm}^2$, we estimate a minimum $1/e$ coherence time for the 4-site chain of $\approx 60$\,\unit{\milli\second}. For the 8-site SSH, we use a detuning of 184.989~GHz, with a shorter estimated $1/e$ coherence time of $\approx 35$\,\unit{\milli\second}. These are worst case estimates, considering the state combination with the largest differential light shifts. The lightshifts could be reduced further by minimising the intensity sampled by each molecule, using a colder molecular gas in a tighter trap or in an optical lattice. Coherence times between 10 states of about 1\,s have been predicted for RbCs molecules in optical tweezers\cite{hepworthLonglivedMultilevelCoherences2025}. The total microwave interrogation time for the 4-site and 8-site spectroscopy measurements was 2\,ms and 8\,ms respectively, significantly shorter than these coherence times. 

The measurements in Figs.\,\ref{fig:SSH_time}\,-\,\ref{fig:SSH_topology} benefit from a long coherence time specifically between the states at the ends of the chain. We therefore set the trap detuning to maximise the coherence time for that pair of states. For the measurements shown in Figs.\,\ref{fig:SSH_time} and~\ref{fig:SSH_topology}, the trap detuning was set to be magic for the states $\ket{1}$ and $\ket{4}$, which we determined to be 184.93(4)\,GHz, by measuring the contrast in Ramsey experiments as in \cite{gregorySecondscaleRotationalCoherence2024}. For these experiments with the 4-site chain, we expect a coherence time for the edge states exceeding $ 300$\,\unit{\milli\second}, limited by the uncertainty with which we have measured the magic wavelength. 

To ensure consistency in the measurements in Fig. \ref{fig:SSH_N_levels}, we set the trap detuning to be magic for the $\ket{1}$ and $\ket{8}$ state combination, even when probing the shorter chains. We measure the magic condition between the $\ket{1}$ and $\ket{8}$ states to be at a detuning of 184.53(1)\,\unit{\giga\hertz}, for which we expect a coherence time of $> 1$\,\unit{\second}. Fitting the results presented in Fig. \ref{fig:SSH_N_levels}, we extract $1/e$ damping times for the edge-state tunnelling of 0.09(6)\,s, 0.09(2)\,s, and 0.9(5)\,s for the 4-, 6-, and 8-site cases, respectively. 



\subsection*{Microwave control and calibrations}

All the microwaves used in our experiment are generated by a single microwave source based on a direct digital synthesiser with a 9\,GHz sampling frequency (AMD RFSoC 4x2 controlled with QICK firmware \cite{stefanazziQICKQuantumInstrumentation2022}). This hardware enables fast frequency switching with ns-scale latency. To maximise microwave power on the molecules, we utilise the two channels of the RFSoC for different microwave frequency ranges. The first channel is used for microwave frequencies below 4\,GHz, while the second channel is operated in the mix-mode configuration to address frequencies above the Nyquist frequency of 4.925\,GHz. The output from each channel is amplified and then combined onto a single microwave antenna placed just outside of the ultra-high vacuum glass cell in which the experiments are performed. For the measurements presented in Figs.\,\ref{fig:SSH_spectroscopy}\,-\,\ref{fig:SSH_topology}, each microwave pulse is $\tau_\text{pulse}=1$\,\unit{\micro\second} long. This is significantly shorter than the lattice tunnelling times $\tau_1$ and $\tau_2$ (equivalently the $\pi$-times of the Rabi couplings) as $\tau_\text{pulse}\simeq \frac{\tau_1}{200} \simeq \frac{\tau_2}{40}$.  For measurements in Fig. \ref{fig:SSH_winding}, as the dynamics occur on shorter timescales, the pulses were set to be 200\,ns long. We measure the rise time of the pulses to be 8\,\unit{\nano\second}, and hence assume that the pulses used in our stroboscopic approach are square in our analysis.


For the 4-site chain, we are limited to Rabi frequencies $<15$\,kHz due to off-resonant coupling to other nearby hyperfine states~\cite{blackmoreCoherentManipulationInternal2020}; transitions to non-stretched states are allowed in our apparatus as the microwaves are unpolarised. However, the Rabi frequencies in the 8-site chain are limited by the radiation efficiency of the antenna used at $4.901$\,GHz, the frequency of the $N=4$ to $N=5$ transition. In this case, the maximum Rabi frequency we are able to reach with continuous-wave driving is 10\,kHz, or $2.5$\,kHz in a stroboscopic sequence with a 25\% duty cycle. 


Our experiments are sensitive to microwave detunings on the scale $\simeq$~50~Hz. We therefore perform spectroscopy of all the transitions in the magic wavelength trap to a precision of $\leq$~10~Hz as part of our calibration routine. We also calibrate the relationship between the microwave power and the Rabi frequencies for each transition by driving Rabi oscillations between individual pairs of states.

\subsection*{Data acquisition and normalisation}

For Fig. \ref{fig:SSH_spectroscopy}, the data used are the average of two experimental runs, taken in blocks with $\Delta_\text{p}$ varied randomly for each fixed value of $J_2$. The data are normalised by the molecule number in the cases with the largest detuning for each block. The data for the other figures are the average of at least three experimental runs. The data here are normalised by the total molecule number present in the sequence. When measuring the population in different states, a small correction factor ($<1.05$) is used to account for the finite fidelity of the $\pi$-pulses. 

\subsection*{Analysis of Trotter Errors}

In the experiment, the target SSH Hamiltonian, $\mathcal{H}_{\text{SSH}}$, is implemented stroboscopically with each sequence implementing a single microwave pulse, $\mathcal{H}_i = J_{i,i+1}(\ket{i}\bra{i+1} + \ket{i+1}\bra{i})$. This accrues Trotter errors of the order of $\mathcal{O}[(J_w\tau)^2]$, where $\tau$ is the pulse time. This can be seen by decomposing the unitary evolution as $U=\prod_{i=1}^{N-1}U_i$ where $U_i=e^{-i\mathcal{H}_i\tau}$. It is straightforward to confirm that $U=e^{-i\mathcal{H}_{\text{SSH}}\tau + \frac{\tau^2}{2}\sum_{i=1}^{N-1} [\mathcal{H}_i,\mathcal{H}_{i+1}] + \mathcal{O}[(J_w\tau)^3]}$. We calculate the Trotter infidelity, $1-\mathcal{F}_{\beta}\approx 0.5\times 10^{-5}$ for the $4$-site chain, where $\mathcal{F}_{\beta} = |\braket{\beta_{\exp}}{\beta_{\text{SSH}}}|^2$ and $\beta_{\text{exp}}$ ($\beta_{\text{SSH}}$) labels the eigenstate of the effective experimental Hamiltonian (actual SSH Hamiltonian).

\subsection*{Theoretical model of the SSH Hamiltonian}

We compare the behaviour of our system against exact diagonalisation of the target SSH Hamiltonian assuming continuous-wave driving fields. We fix the Rabi frequencies to the values we independently measure by driving each transition in isolation, and set the detunings to zero for on-resonance driving, or at the experimentally measured value when non-zero in Fig.\,\ref{fig:SSH_topology}(b). We propagate the uncertainties on the Rabi frequencies and detunings through this model via a Monte Carlo approach. We assume a Gaussian distribution of the errors and perform multiple simulations with parameters drawn from these distributions, from which a 1-$\sigma$ error is calculated. 


\section*{Data availability}
All data presented in this work are available at doi:10.15128/r1pr76f3548

\section*{Acknowledgements}
We thank Tom Hepworth for insightful discussions on the magic wavelength optical trapping and for helping to develop the control of the RFSoC device. We acknowledge support from the UK Engineering and Physical Sciences Research Council (EPSRC) Grants EP/P01058X/1, EP/W00299X/1, EP/W016141/1, EP/Y01510X/1 and UKRI2226 funded through the Programme Grant Scheme, UK Research and Innovation (UKRI) Frontier Research Grant EP/X023354/1, the Royal Society, and Durham University. PDG is supported by a Royal Society University Research Fellowship URF/R1/231274 and HP by a Royal Society University Research Fellowship UF160112. K.H.'s work was supported in part by the National Science Foundation (DGE-2346014) and the W. M. Keck Foundation (Grant No. 995764).

\bibliography{refs,extrareferences}


\setcounter{equation}{0}
\setcounter{figure}{0}
\setcounter{table}{0}
\setcounter{page}{1}
\makeatletter
\renewcommand{\theequation}{S\arabic{equation}}
\renewcommand{\thefigure}{S\arabic{figure}}
\renewcommand{\bibnumfmt}[1]{[S#1]}
\renewcommand{\citenumfont}[1]{S#1}

\end{document}